\documentstyle[aps,prc,preprint,tighten]{revtex} 
\begin{document} 
\preprint{KSUCNR-102-98} 
\draft 

\def\rightmark{Saturation Properties of Nuclear Matter}
\def\leftmark{C. W. Johnson and G. Fai}

\vskip .4 cm
 
\title{Saturation Properties of Nuclear Matter\\
with Nonlocal Confining Solitons$^{*}$ 
}
\author{
{Charles W. Johnson$^{ \dag}$ and George Fai$^{ \ddag}$
}\\[2.812mm]
{\normalsize
Center for Nuclear Research\\
Department of Physics\\
Kent State University\\
Kent, OH 44242 USA\\[0.2ex]
}}
\maketitle
 
\begin{abstract}
We examine saturation properties of a quark-based picture of nuclear matter.
Soliton matter consisting of nonlocal confining solitons is used to model nuclear
matter. Each composite nucleon is described by a non-topological soliton as given by 
the Global Color Model. We apply techniques and concepts from the description of 
crystal lattices. In particular, the Wigner-Seitz approximation is used to calculate the 
properties of the soliton lattice at the mean-field level. We focus on infinite
nuclear matter at around standard nuclear matter density with the simplest one-parameter
assumption for the gluon propagator. The saturation density and incompressibility 
are calculated as functions of the single input parameter of the model.
\end{abstract}

\section{Introduction}
The Global Color Model (GCM)\footnote{Not to be confused with the ``Generator 
Coordinate Method'' with the same acronym.} 
has been developed to model quantum chromodynamics (QCD) 
at energies low on the scale of high-energy particle physics\cite{cahill85}.
The successes of the GCM in the hadronic sector\cite{tandy97} include
the reproduction of chiral perturbation theory results
\cite{cahill85,roberts94;a,frank96;a}, meson form factors\cite{frank95} and spectra\cite{frank96;b},
and both, the soliton\cite{frank91} and the
Faddeev\cite{cahill92} picture of the nucleon. These successes motivate
the application of this particular soliton model when
a description of nuclear matter in terms of quarks and gluons is attempted. 

Vladimir Naumovich Gribov efficiently used condensed-matter techniques in studies 
of quark confinement\cite{gribov97}. This is but one example of the recent convergence 
in language and methods between nuclear/particle physics and 
condensed-matter physics. In this paper we use the theory 
of crystal lattices in a description of 
nuclear matter in terms of quarks and gluons, with the GCM providing the 
physics of the primitive cell, which corresponds (for large values of the lattice constant) 
to an isolated composite nucleon.

To model nuclear matter in terms of a many-soliton problem where each individual
soliton is accountable to the underlying quark and gluon degrees of freedom is an 
ambitious undertaking\cite{banerjee85,reinhardt85,achtzehnter85}.
At low excitation the kinetic energy of the composite 
solitons can be neglected as a simplification. Due to the short-range repulsion
between the solitons, a soliton crystal emerges as the lowest-energy solution at
the mean-field level\cite{cohen89}. This crystal, however, is complicated
relative to the structures familiar from condensed matter physics in that the 
lattice sites are themselves populated by extended objects (solitons), localized to
some extent for large lattice constants. As the density increases, we expect a 
mean-field solution which no longer has localized centers of density.  

An early success of condensed-matter physics is the Kronig-Penney model\cite{kronig31},
which treats a crystal as a one-dimensional periodic set of square wells. This simple description 
allows analytic solutions to be found for the Bloch wave functions and for the energy bands 
usually associated with crystal lattices. In the case of nuclear matter, deeply bound light quarks 
call for a relativistic approach. Glendenning and Banerjee have solved the relativistic 
Kronig-Penney model and obtained analytic solutions for the eigen-value 
spectrum which retains the band structure\cite{glendenning86}.

In more complete geometries analytic solutions of the Dirac equation for the Bloch wave functions
and eigen-energies are not available, so some kind of approximation is needed.
Here we apply the Wigner-Seitz approximation\cite{wigner33} to nuclear matter in the ground state. 
This entails the assumption of a spherically symmetric mean field, which 
takes into account the effect of the surrounding matter (lattice sites)
on the primitive cell in a directionally averaged manner. This approximation may 
quite naturally take into account that nuclear matter is more like a fluid than a
crystal\cite{birse88}.

As discussed above, the primitive cell will be given by the GCM, which
admits soliton solutions with an intrinsically generated, extended $\bar{q}q$ meson field. 
In addition, the individual GCM solitons are confining, as can be seen by the lack of 
poles in the quark propagator outside the region where the meson field is 
nonzero\cite{frank91,jff96}. We have recently applied this kind of picture to draw attention 
to the intersection of the ground state band with the next unoccupied band 
at high densities\cite{jff96,jf97}. The intersection represents a color insulator-to-conductor
transition in the model, and thus signals the onset of quark deconfinement. In this sense 
these ideas of condensed matter physics are applied in the spirit
of Gribov\cite{gribov97}, albeit to a model situation in place of full QCD.   

Our goal in the present paper is to look at the saturation properties of this description
more closely. The paper is organized as follows. In Section 2, we describe the model used 
for the primitive cell. The application of the GCM to nuclear matter is reviewed in Section 3. 
In Section 4 we present new results around saturation density, including 
nuclear-matter incompressibility. Section 5 contains the discussion and the summary. 

\section{GCM: the primitive cell}\label{gcm}

As quarks and gluons are confined within hadrons, the 
description of low-energy nuclear matter has traditionally been accomplished in terms of 
effective hadronic degrees of freedom. The effective degrees of 
freedom should be derived from the QCD Lagrangian. However, until an
appropriate solution of QCD at low energies and large length-scales becomes available,
it is necessary to use a model in place of full QCD. Clearly, we attempt
to keep as much of the essential physics in the model as possible. 
An approach to systematize the needed approximations and to 
connect QCD to an effective hadronic field theory can be formulated in 
terms of functional integral methods. The strategy is to 
transform the integration variables from quark and gluon fields to hadron 
fields.

One particular implementation of the above ideas is in the framework
of the GCM, which starts with a truncation 
of QCD\cite{cahill85}, leading to the Euclidean action$^{\S}$
\begin{equation}
S[\bar{q}q]=-\int d^4 x~d^4 y \left[ \bar{q}(x)(\gamma \cdot \partial +m) 
\delta (x-y)q(y)+\frac{g^2}{2}j^a_{\mu}(x)D_{\mu \nu}(x-y)j^a_\nu (y) \right].
\label{eq:action}
\end{equation}
Here $j^a_ \nu (x) = \bar{q}(x)\frac{ \lambda_a}{2} \gamma_ \nu q(x)$ is a 
local quark current, with Euclidean
Dirac matrices $\gamma_{\nu}$ and Gell-Mann matrices 
$\lambda_a$. The two-point gluon function, $D_{\mu\nu}$, can be 
considered the phenomenological input point for the model. 
Using a Feynman-like gauge, $D_{\mu\nu}=\delta_{\mu \nu}D(x-y)$,
the gluon propagator is particularly simple, and provides a single
parameter function for 
the GCM. In (\ref{eq:action}), $m$ is a current quark mass, which will be taken to be zero 
in the following. The GCM has the global symmetries of QCD, but lacks local gauge invariance.
Our choice of the gluon propagator is dictated by the requirements of simplicity and 
quark confinement. We take a delta function in momentum space,  
\begin{equation}
g^2 D(q)=3\pi^4 \alpha^2 \delta^{(4)} (q) \,\, .   
\label{gluon}
\end{equation}
The strength parameter $\alpha$ is the only input of the model. 
This form of the gluon propagator has been shown to produce quark confinement 
and reproduce nucleon properties\cite{frank92}.
 
To exhibit nonlocal quark-antiquark structures in the action, 
a Fierz reordering may be performed\cite{cahill88}, which
transforms the current-current term in~(\ref{eq:action}) as
\begin{equation}
\frac{1}{2} \int d^4 x~d^4 y~j^a_\mu(x) D(x-y) j^a_\mu(y) =
-\frac{1}{2} \int d^4 x 
~d^4 y~ {\cal J}^{\theta}(x,y) D(x-y) {\cal J}^{\theta}(y,x) \,\, .  
\end{equation}
Here, ${\cal J}^{\theta}(x,y)=\bar{q}(x)\Lambda^{\theta} q(y)$ can be looked
upon as a quark-antiquark bilocal current with quantum numbers specified by $\theta$.
The quantity $\Lambda^{\theta}$ is a direct product of Dirac, flavor, and 
color matrices, and contains both, color-singlet and 
color-octet terms. We focus on the 
color-singlet sector in this work, ignoring 
correlations that correspond to diquark degrees of freedom.

To cast the partition function in terms of Bose fields, 
auxiliary nonlocal fields, ${\cal B}^{ \theta}(x,y)$, are introduced,
and the partition function is multiplied by 
\begin{equation}
1=N\int {\cal D} {\cal B} \exp \left[ - \int 
d^4 x d^4 y \frac{ {\cal B}^{ \theta}(x,y){\cal B}^{ \theta}(y,x)}{2g^2 D(x-y)}
\right]  \,\,\, ,
\end{equation}
where $N$ is a normalization constant.
After the transformation ${\cal B}^\theta (x,y) \rightarrow 
{\cal B}^\theta (x,y)+g^2 D(x-y){\cal J}^{ \theta} (y,x)$, the action is 
bilinear in terms of the quark fields and the Grassman integration can be 
performed in the usual way. This yields the action in terms of bilocal Bose fields.  

The replacement of the quark fields with Bose fields is, in principle,
an exact functional change of variables. Observables calculated from the 
action (\ref{eq:action}) are not affected by the variable transformation, but are 
now expressed in terms of the Bose degrees of freedom, provided the entire sum
over $\theta$ is kept. This is impossible in practice, and the truncation
scheme used can be developed into a systematic method of approximation.
To retain the chiral content of the QCD action, at least two Bose fields
need to be kept. 

The classical vacuum configuration ${\cal B}_0^{ \theta}$ is identified by
$\delta S/ \delta {\cal B}^{ \theta} = 0$. 
This produces a quark self-energy, 
$\Sigma(x-y)= \Lambda^{ \theta} {\cal B}_0^{ \theta} (x-y)$ 
satisfying a Schwinger-Dyson equation. In momentum space 
\begin{equation}
\Sigma(p)=i \gamma \cdot p[A(p^2)-1]+B(p^2)=g^2 \int \frac{d^4 q}{(2 \pi 
)^4 } D(p-q) \frac{ \lambda^a }{2} \gamma_{ \mu} \frac{1}{i \gamma \cdot q +m+ 
\Sigma(q)} \frac{ \lambda^a}{2} \gamma_{ \mu} \,\,\, .
\label{DS}
\end{equation}
This can be considered an integral equation for the self-energy 
amplitudes $A(p^2)$ and $B(p^2)$. 
Numerical solutions for these amplitudes are now 
available at varying degrees of sophistication\cite{roberts94;b}. Our choice for this study
is governed by simplicity within the context of the requirement of confinement.
The delta-function gluon propagator (\ref{gluon}) simplifies the integral equation 
(\ref{DS}) resulting in an algebraic equation. 
Analytical solutions for $A$ and $B$ can be found as\cite{munczek83}
\begin{eqnarray}
A(p^2) = \left\{ \begin{array}{c} 2 \\
	 \frac{1}{2}[ 1 + (1+\frac{2\alpha^2}{p^2})^{\frac{1}{2}}]
		     \end{array}  \right. \;\; ,  & 
B(p^2) = \left\{ \begin{array}{c} (\alpha^2-4p^2)^{\frac{1}{2}} \\
	                               0
		     \end{array}  \right. \;\; & 
		 \begin{array}{c} p^2 \leq \frac{\alpha^2}{4} \\
				  p^2   >  \frac{\alpha^2}{4}	 		
		     \end{array}	\;\; ,    
\label{AB}
\end{eqnarray}
where the input parameter $\alpha$ controls the coordinate space width of $A$ and $B$. 
The lack of solutions to the equation $p^2+M^2(p^2)=0$, (where $M=B/A$ is the dynamic quark mass) 
indicates that (\ref{AB}) produces quark confinement as there 
is no on-mass-shell point, thus the propagation of a quark in the normal vacuum
is prohibited\cite{jff96}. It is important to note that the amplitude $B(p^2)$
plays a dual role in the model: it also acts as the distributed vertex 
for coupling the quarks to the $\bar{q}q$ Goldstone modes\cite{frank91,delbourgo79}.

The fluctuations $ \widehat{ {\cal B}}^{ \theta}(x,y)= 
{\cal B}^{ \theta} (x-y)- {\cal B}^{ \theta}_0$ are identified 
as the propagating Bose fields. If the 
color-singlet scalar-isoscalar and pseudoscalar-isovector fluctuations 
are retained, the formalism can be adopted to the requirement
of chiral invariance by the variable transformation
\begin{equation}
\Lambda^{ \theta} \widehat{ {\cal B}}^{ \theta}(x,y)=
\frac{B(r)}{f_{ \pi}} \widehat{ 
\chi}(R)e^{ i \gamma_5 \cdot \phi(R)/f_ \pi} \,\,  ,
\label{chsym}
\end{equation}
where $r=x-y$ and $R=(x+y)/2$, are relative and cm-like coordinates, 
respectively, $f_{\pi}$ is the pion decay constant,
and it has been assumed that
the on-shell form factor $B$ can also be used off-shell. 
As a further simplification, the
$\phi=0$ point on the chiral circle can be fixed. In this case the
radial fluctuations away from the chiral circle coincide with
the scalar-isoscalar field variable prior to the transformation.
In the numerical work that follows the single fluctuation field 
$\widehat{\chi}$ corresponding to this situation will be used.      
Letting $m \rightarrow 0$, 
the action (up to a constant) can be written as a sum of fermionic
and bosonic terms:
\begin{equation}
S[ \mu, \widehat{ \chi}]=-Tr[~ \ln G^{-1} ( \mu, \widehat{ \chi})- \ln G^{-1}
(0, \widehat{ \chi})]+ \int d^4 R~[ \frac{1}{2}( \partial_{ \mu} \widehat{ 
\chi})^2 + U( \widehat{ \chi}^{2})]~.
\label{actfb}
\end{equation}
The chemical potential ($\mu$) dependence of the fermion term 
in equation (\ref{actfb}) ensures 
that a meson source from the valence quarks will be generated\cite{frank91}. 
The $U( \widehat{ \chi}^2)$ term is the effective meson 
self-interaction\cite{cahill85}. For $\mu=0$ the inverse quark propagator 
takes the form
\begin{equation}
G^{-1}(x,y)= \gamma \cdot \partial_x~ A(x-y)+f^{-1}_{ \pi} B(x-y) 
\widehat{ \chi} \left(\frac{x+y}{2} \right)~,
\end{equation}
and the saddle-point configuration is at $\widehat{ \chi}=f_{ \pi}$\cite{frank92}.

Since $G^{-1}(x,y)$ is time independent, stationary 
eigenstates of the form $u_j ({\bf x})$ can be obtained from a
self-consistent Dirac equation, which in coordinate space takes the form
\begin{equation}
0=\int d^3 y \left\{(- \gamma_4 \epsilon_j + \vec{\gamma} \cdot \nabla 
)A({\bf x}-{\bf y})+B({\bf x}-{\bf y})+\frac{B({\bf x}-{\bf y})}{f_{ \pi}} 
\chi \left( \frac{ {\bf x}+{\bf y}}{2} \right) \right \}u_j({\bf y})~.
\label{dirac}
\end{equation} 
The new $\chi= \widehat{ \chi}-f_{ \pi}$, and as $p_4 = q_4 = i\epsilon_j$,
where $\epsilon_j$ is the energy eigenvalue, the meson vertex $B$ has an 
energy dependence. It can also be seen that 
a wave-function renormalization appears with the renormalization
constant $Z_j$ given by\cite{frank91}
\begin{equation}
Z_j=- \int d^3 p~ d^3 q~ \bar{u}_j ( {\bf p} ) \frac{ \partial G^{-1} (i 
\epsilon ; {\bf p}, {\bf q} ) }{ \partial \epsilon _j} u_j( {\bf q})~.
\end{equation}
The meson field equation $\frac{ \delta E}{ \delta \chi}=0$ may be 
summarized as
\begin{equation}
- \nabla \chi ({\bf z})+ \frac{ \delta U}{ \delta \chi ({\bf z})} + 
Q_{ \chi}({\bf z})=0 \,\, ,
\label{KG} 
\end{equation}
with the meson source provided by the valence quarks according to
\begin{equation}
Q_{ \chi} ({\bf z})= \sum_{j} \frac{1}{f_{ \pi} Z_j} \int d^3 x~d^3 y~ 
\bar{u}_j ( 
{\bf x}) B(- \epsilon^2_j ; {\bf x} - {\bf y} ) \delta \left[ \frac{ {\bf x} + 
{\bf y} }{2} -{\bf z} \right] u_j( {\bf y} )  \,\, .  
\label{mesource}
\end{equation}

Equations (\ref{dirac}) and (\ref{KG}) form a system of coupled differential
equations for the quark wave functions and the meson field, which need to be
solved selfconsistently, with the appropriate boundary conditions.

\section{Infinite Nuclear Matter}\label{inm}

The use of soliton matter to model high-density nuclear matter originated in the  
mid-eighties\cite{banerjee85,reinhardt85,achtzehnter85}. The new feature of the present 
work is that the GCM soliton represents the primitive cell. 
Some of the above authors also 
apply the Wigner-Seitz approximation\cite{wigner33}, to which we turn next.

\begin{figure}[t]
\vspace*{7.0cm}
\includegraphics{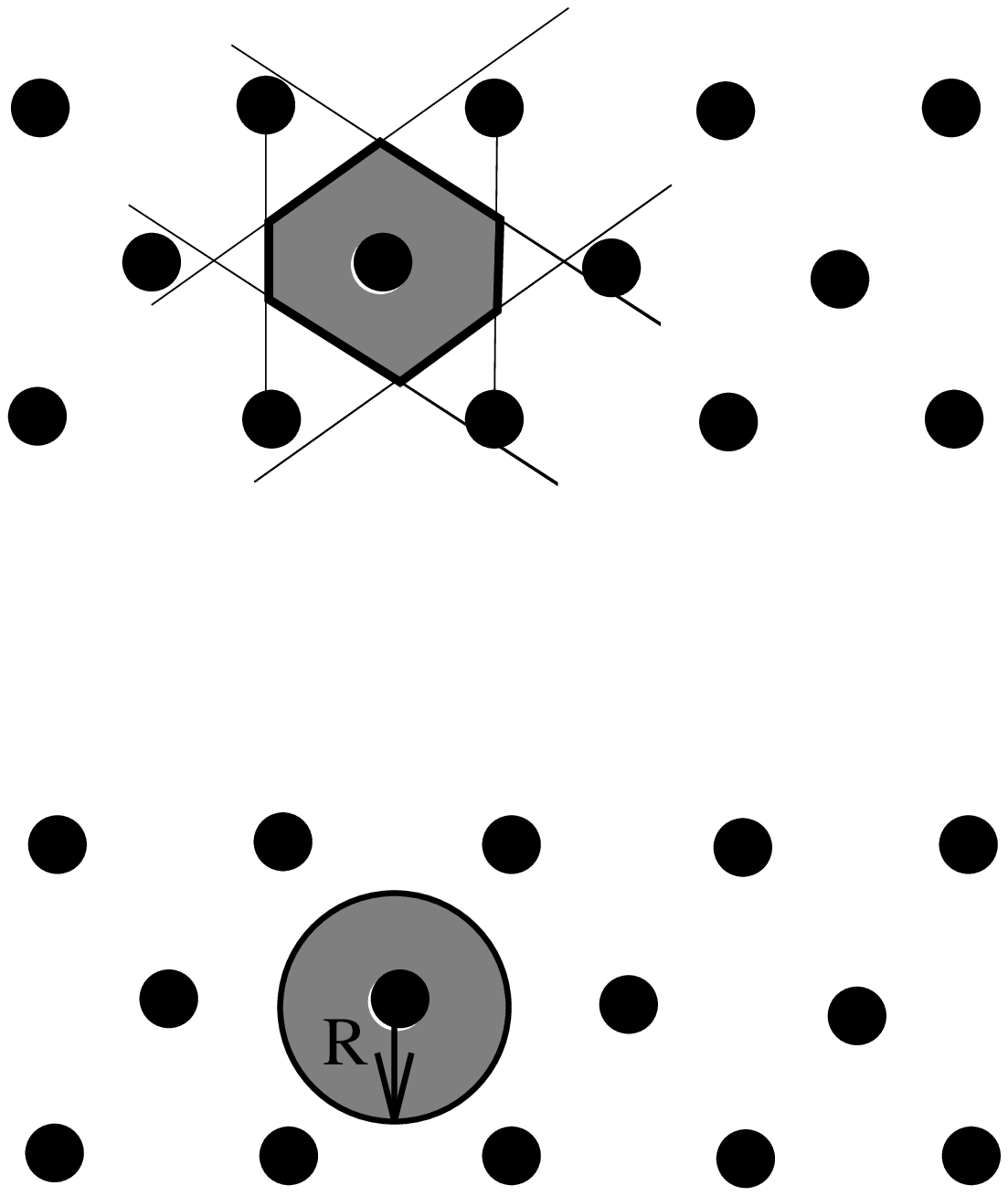}
\caption[]{The top figure shows what the primitive cell looks
like for a simplistic two dimensional case. The bottom figure shows what the
primitive cell looks like in the Wigner-Seitz approximation.}
\label{seits_cellps}
\end{figure}

\subsection{Wigner-Seitz approximation}

As a means of describing nuclear matter, we consider an infinite 
collection of GCM solitons.  At the lowest energies 
the solitons are expected to arrange themselves in a crystal 
lattice\cite{glendenning86,cohen89}. Accordingly, the single-quark 
eigen-energies will develop into energy bands. 
For simplicity, we assume a simple cubic crystal (sc). 
For a periodic lattice, the Dirac wave function must 
be invariant to a lattice translation, so the solutions must have 
the Bloch form~\cite{kittel86}
\begin{equation}
u^{lat}_{{\bf m}} ({\bf r})=
u_{ {\bf m} } ( {\bf r}) \,\, e^{ {\textstyle i {\bf m} \cdot {\bf r }}} \,\, ,
\label{bloch}
\end{equation}
where ${\bf m}=(m_x ,m_y ,m_z )$ is the lattice momentum and $u_{ {\bf m} }
( {\bf r})$ 
is a Dirac spinor which has the periodicity of the lattice. 
To solve for the 
Bloch functions we employ the Wigner-Seitz approximation~\cite{wigner33}.
This amounts to taking the primitive cell as given by the geometry of the lattice
(a cell bounded by perpendicular bisectors of the lattice vectors
as shown in Figure~\ref{seits_cellps}) and replacing it by a spherical cell of
radius $R$ and solving for ${\bf m}=0$ in (\ref{bloch}). The full solution
$u^{lat}_{{\bf m}} ({\bf r})$ is then approximated by 
$u_0 ( {\bf r}) \,\, e^{ {\textstyle i {\bf m} \cdot {\bf r }}}$.  
Changing the density will be implemented in our calculation 
by varying the cell radius $R$. 

The Wigner-Seitz approximation places specific boundary conditions on the 
Dirac spinors. These conditions express the requirement that the upper
component of the wave function must be periodic and anti-periodic for the 
bottom and the top of the band, respectively. We focus attention on 
the lowest-energy state of the band, for which the above, together with 
the $r=0$ boundary conditions, implies 
\begin{equation}
   \left. \left.  g'(r) \right|_{r=R} = f(r) \right|_{r=R} = 0   \,\, ,
\label{bc}
\end{equation}
where $g(r)$ and $f(r)$ represent
the radial parts of the upper and lower components of the Dirac wave 
function, respectively. In addition, 
the meson field of equation (\ref{KG}), which also appears in the source 
term of the Dirac equation (\ref{dirac}) 
must now be periodic in $r$, so that 
\begin{equation}
\left. \chi(r+2R)=\chi(r)~; \,\,\chi'(r) \right|_{r=R}=0 \,\,  ,
\label{sig_bc}
\end{equation}
where $\chi(r)$ is the radial part of the meson field.

\subsection{GCM on the Lattice}

It is convenient to solve the Dirac equation in momentum space, while the 
nonlinear Klein-Gordon equation is easier to handle in coordinate space.
We seek solutions  of the Dirac equation with the boundary conditions (\ref{bc}). 
These can be incorporated using a three dimensional Fourier expansion. Expanding 
$A$, $B$, $u$, and the meson-field source, we integrate over $y$ and use 
orthonormality to get an equation for the Fourier components of the 
quark wave function 
\begin{displaymath}
\left\{ \left[
\begin{array}{cc}
 (-\epsilon A(k_n)+B(k_n))\delta_{nm} &-k_nA(k_n)\delta_{nm}\\
 k_nA(k_n)\delta_{nm}& (\epsilon A(k_n)+B(k_n))\delta_{nm} 
\end{array}
\right]\right.+
\end{displaymath}
\begin{equation}
4\pi
\left. \left[
\begin{array}{cc}
{\displaystyle \sum_{m=0}^{\infty}}k_m^2 V_0(k_n,k_m) & 0\\
0 & {\displaystyle \sum_{m=0}^{\infty}}k_m^2V_1(k_n,k_m)
\end{array}
\right]\right\}\left[
\begin{array}{l}
 g_m \\  f_m
\end{array}
\right]
 = 0 \,\, ,
\label{lat_dirac}
\end{equation}

where $g_n=g(k_n)$, $f_n=f(k_n)$, and 

\begin{equation}
V_l=\int_{-1}^{1}B\left(\frac{{\bf k}_n + {\bf k}_m}{2}\right)
\chi({\bf k}_n- {\bf k}_m)
P_l(\cos\theta)d(\cos\theta)~.
\label{v_l}
\end{equation}

To get this final form we have written the Dirac wave functions as
\begin{equation}
u({\bf k}_n )=\left[ \begin{array}{c}
                       g(k_n)\\ i {\bf \sigma } \cdot \hat{k}_n f(k_n)
                      \end{array}
               \right] {\cal Y}^{m_j}_{jl}( \hat{k}_n)   \,\, ,
\label{sphere}
\end{equation}
where $\displaystyle k_n=\frac{n \pi}{R}$ and we used the spherical symmetry 
of the Wigner-Seitz cell.
If $m = 1,2, \ldots, M$, then
equation (\ref{lat_dirac}) is a $2M$-by-$2M$ eigenvalue problem for 
the energy eigenvalue $\epsilon$. The quantity $B/A$
plays the role of a dynamic mass and 
the scalar part of the self-energy $B$ also acts to
couple the quarks to the meson field via~(\ref{v_l}). The 
self-energy terms have an $\epsilon$ dependence which makes this a highly 
non-linear problem. The $V_l$ term ($l=1,2$) represents
the Legendre coefficient for the meson field in the presence of the  
distributed coupling $B$. One needs to solve the Dirac equation~(\ref{lat_dirac}) 
and the Klein-Gordon equation for the meson field (\ref{KG}) selfconsistently. 

To solve for a soliton lattice, we 
first pick a starting meson field and search for the lowest energy eigenvalue 
of  Eq. (\ref{lat_dirac}). This means finding the energy $\epsilon$ 
which makes the determinant of equation (\ref{lat_dirac}) vanish.
We start at $\epsilon=0$ 
and work upwards in energy until the determinant changes sign. We then use 
the bisection method to find the root. Care must be taken so that the 
initial steps are sufficiently fine in $\epsilon$ not to miss the lowest root. 
With the root in hand, we can solve for the 
Fourier components of the Dirac wave functions. For this we first perform a 
lower-upper triangular decomposition and use inverse 
iteration\cite{press92}. The
momentum-space meson-field source term is constructed from the Dirac wave 
functions and we transform the source to coordinate space for use in the 
nonlinear Klein-Gordon equation for the meson field (\ref{KG}). 
To solve this nonlinear equation, we treat equation (\ref{KG}) as a functional 
of $\chi$ and use Newton's method. 
Once the Klein-Gordon equation is solved
for the new meson field, we start 
over with the Dirac equation in this modified meson field.
We iterate until convergence of the quark wave functions is achieved, which 
takes between three to six iterations to reasonable accuracy.
 
\subsection{Energy Bands}

As known from condensed-matter physics, each soliton of the lattice contributes 
one level to each energy band\cite{kittel86}.
In the Wigner-Seitz approximation we need to calculate only the energy for 
the case when the crystal momentum ${\bf m}$ in equation (\ref{bloch}) is zero. 
The Dirac-Bloch wave function, in the Wigner-Seitz approximation, for an arbitrary 
crystal momentum is
\begin{equation}
u^{lat}_{{\bf m}} ({\bf r})=u_{0} ( {\bf r})~e^{ {\textstyle i {\bf m} \cdot 
{\bf r }}} \,\, ,
\label{bloch_0}
\end{equation}
and we use (\ref{bloch_0}) to calculate the expectation value for the 
square of the Dirac Hamiltonian to estimate the lattice 
momentum dependence of the energy levels in the band as
\begin{equation}
\epsilon_m=[ \epsilon_{bot}^{2}+{\bf m}^2]^{\frac{1}{2}}  \,\, ,
\label{en_band}
\end{equation}
where $\epsilon_{bot}$ is the energy of the bottom band.

To obtain the possible values of ${\bf m}$ we refer back to the underlying lattice structure. 
For a simple cubic crystal of $N$ solitons and
sides of length $L=2RN$, 
the allowed values of the component of the lattice momentum 
in the direction of any of the three axes are 
\begin{equation}
m=0,~ \pm \frac{2 \pi}{L},~\pm\frac{4\pi}{L} \ldots,\frac{N\pi}{L}~ \,\,  ,
\label{latt_mo}
\end{equation}
with the top of the energy band corresponding to $\displaystyle m=\frac{N \pi}{L}=\frac{ 
\pi}{2R}$. Thus for
the top energy band we obtain
\begin{equation}
\epsilon_{top} = \left[ \epsilon_{bot}^{2}+( \frac{ \pi}{2R})^2  \right]^{\frac{1}{2}}~.
\label{top_band}
\end{equation}

We have performed the same estimate assuming body-centered and face-centered
cubic lattices. This variation in assumed crystal structure introduces a 
less than 10 \% uncertainty in our results.  

The lowest band ($l=0$, $l'=1$ and $j=\frac{1}{2}$) is labeled 
$1s_{1/2}$. The next lowest (at around standard nuclear density)
band has nonzero orbital angular momentum
in the large component of the Dirac wave function
($l=1$, $l'=2$, and $j=\frac{3}{2}$) and is labeled $1p_{3/2}$. 
The next band is again an $s$-state, corresponding to a radial
excitation. For very low density ($R \longrightarrow \infty$) the energy bands
shrink to single levels and in the limit reproduce the
energies of a single soliton. As the density increases, the bands spread out 
and approach each other. The intersection of the energy bands as the density
is increased beyond twice standard nuclear density
is discussed elsewhere\cite{jf97}. However, properties
of the model at around saturation density have not been examined in detail 
in Ref. \cite{jf97}. This is the goal of the present study.

\section{Results}\label{results}

Here we focus on the properties of the lowest energy band (ground state band)
at around standard nuclear density, $\rho \approx \rho_0=0.17$ fm$^{-3}$. 
With one soliton in each Wigner-Seitz cell,
the energy of each primitive cell is given by the soliton energy. As mentioned above,
the kinetic energy of the center-of-mass of the soliton is neglected in the present
approximation. The total soliton energy has a quark contribution which contains the
quark kinetic and potential energies. Assuming three-fold degeneracy of the energy levels 
(three colors), the quark energy thus equals three times the lowest
quark eigen-energy $(\epsilon)$ in the self-consistent meson field. 
In addition, the kinetic and potential 
energy associated with the meson field contribute to the soliton energy. 
The meson field has a kinetic energy given by
\begin{equation}
KE_{ \chi}=\int d^3 x \left[ \frac{1}{2} \left( \nabla \chi \right)^2 \right]~~.
\label{ke}
\end{equation}
The potential energy of the meson field is given by
\begin{equation}
PE_{ \chi}=\int d^3 x~U( \chi )~~,
\label{pe}
\end{equation}
where $U( \chi )$ is the meson self-interaction from eq. (\ref{actfb}),
expressed as a function of the fluctuation field $\chi = \widehat{\chi}-f_{\pi}$.
$U(\chi)$ has the usual ``Mexican hat'' shape\cite{frank92}. The total energy takes the form
\begin{equation}
E=3 \epsilon + KE_{ \chi}+PE_{ \chi} ~~.
\label{te}
\end{equation}
This quantity is plotted against the density in terms of standard nuclear density
in Figs. 2-4 for different values of the parameter $\alpha$. We selected a range 
for the values of $\alpha$ earlier\cite{johnson98}. It was argued that while reproducing
the pion decay constant $f_{\pi}$ closely is particularly important for studies at the hadron
level, for the present work, which is concerned with the density of nuclear matter, 
the root mean square (rms) radius of the proton constitutes the most sensible benchmark. It is not
realistic to expect that this simplified one-parameter model, which does not explicitly
take the pion field into account, should simultaneously fit both of these quantities. 
Since the explicit pion cloud is neglected in the model and it is normally assumed that the 
pions will increase the value of the rms proton radius by about 20-30 \%, we need to choose 
$\alpha$ to give an rms proton radius $\approx .6$ or .7 fm. We are able to achieve this for
$\alpha \approx 1.35 \pm .1$ GeV. Thus, we display calculated results for
the values $\alpha = 1.25, 1.35$ and 1.45 GeV  in the following.

In Fig. 2 we show the quark contribution and the total soliton
energy for $\alpha=1.25$ GeV. On one of the calculated points we indicate a typical 
uncertainty we associate with our computation. The main source of this uncertainty, 
which is not more than a couple percent, is the freedom in prescribed tolerances 
at different stages of the calculation. 
We see that as the density increases from zero,
the ground-state energy develops a minimum. The low-density attraction between 
the solitons is a consequence of the boundary conditions on the quark
wave functions~(\ref{bc}).

\begin{figure}[t]
\vspace*{8.0cm}
\includegraphics{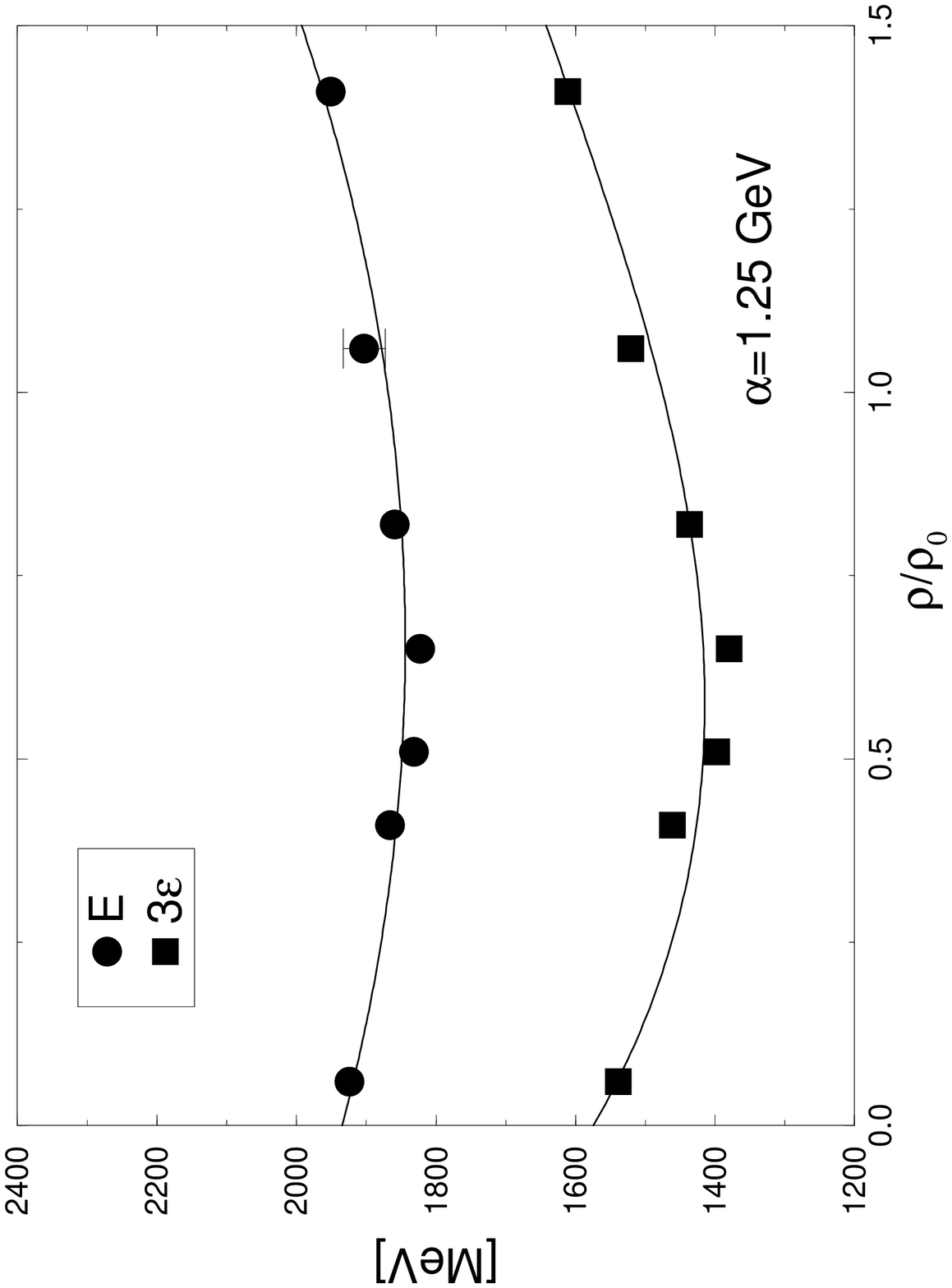}
\caption[]{The total energy $E$ and the quark contribution to the total energy in MeV, 
as a function of $\rho/\rho_0$ for $\alpha=1.25$~GeV.}
\label{fig1}
\end{figure}
 
In particular, the upper component of
the quark wave function is forced to have less curvature than in the 
case of a single soliton, leading to a lower value of the quark kinetic 
energy. At higher densities, where the solitons and the quark wave functions
begin to overlap, the resulting repulsion overcomes the attraction and the
ground-state quark energy starts to increase.
For this value of $\alpha$ there is a minimum in the total energy 
at $\rho_{min} /\rho_0 =0.66$. 
To calculate the minimum we performed a second-order fit to the calculated points around the 
minimum, up to $\rho/ \rho_0 =1.5$. The value of the total energy at the minimum is
$1844$ MeV. However, our scheme does not take into account the center-of-mass kinetic
energy associated with the quarks. An estimate of the soliton mass (on the lattice)
can be obtained by subtracting this motion;  
$M_s = [E^2 - \langle P^2 \rangle ]^{1/2}$, where $\langle P^2 \rangle = 3 \langle p^2 \rangle$,
with $ \langle p^2 \rangle$ being the expectation value of the square of the 
quark momentum. A numerical estimate along these lines leads to a soliton mass
$\approx 15$ \% below the soliton energy, yielding $\approx 1560$ MeV for the soliton mass
for $\alpha = 1.25$ GeV. Scaling back to $\alpha = 1.05$ GeV (used in the earlier 
single-soliton work) would decrease this by another 20 \% to $\approx 1,200$ GeV,
not too far from the average mass of the nucleon-delta system.
Explicitly including the pion will lower this value even further\cite{frank92}.

Figs. 3 and 4 contain the same information for $\alpha=1.35$ and 1.45 GeV, respectively.
In Fig. 3 we observe a minimum of the total energy at $\rho_{min} /\rho_0 =0.73$.  
Fig. 4 yields a minimum at $\rho_{min} /\rho_0 =0.98$.  
We have included all calculated points below $\rho/ \rho_0 =1.5$ in all these fits.
(We carried out calculations at higher densities for each value of $\alpha$ in connection with
our studies of the QGP transition\cite{jf97}. However,
in the present work we focus on 
the second-order behavior of the energy versus density curve near its equilibrium value 
(minimum). We estimate an uncertainty in the position of the minima
on the order of 15\%, as a consequence of combining uncertainties in crystal structure,
tolerances and fits.
 
\begin{figure}
\vspace*{8.0cm}
\includegraphics{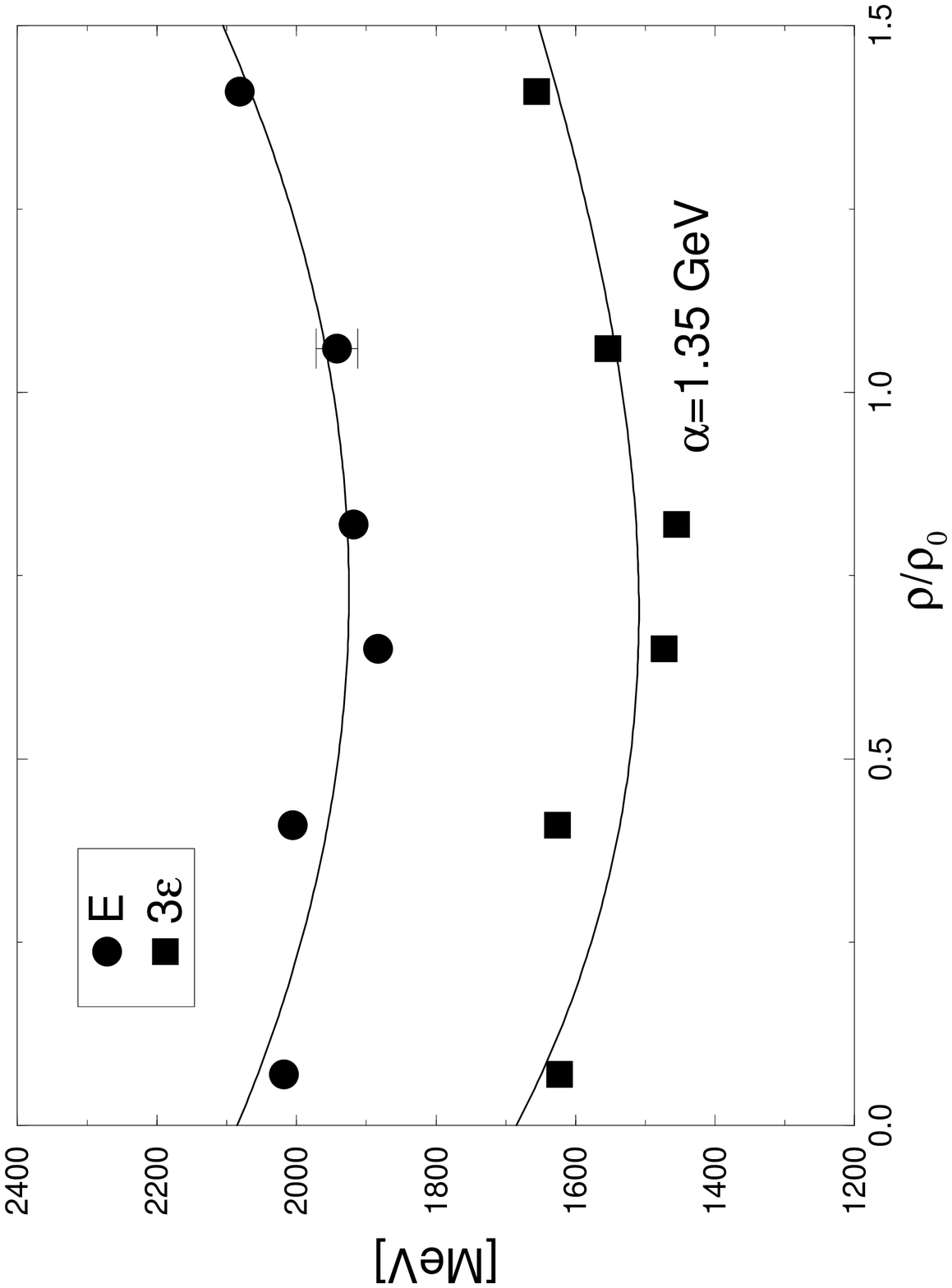}
\caption[]{The total energy $E$ and the quark contribution to the total energy in MeV, 
as a function of $\rho/\rho_0$ for $\alpha=1.35$~GeV.}
\label{fig2}
\end{figure}

Finally we calculate the incompressibility defined as 
\begin{equation}
K ( \rho = \rho_{min})=9 \rho \frac{ \partial E}{ \partial \rho} 
\label{comp}
\end{equation}
around the observed minima. The resulting values are summarized in Table 1,
together with the pion decay constant, the root mean square proton (single soliton) radius and the 
value of the density where the minimum occurs for each value of $\alpha$. As 
$\alpha$ increases, the root mean square radius of a single soliton is moving away from
the experimental value of the rms radius of the proton,
$\displaystyle \langle r^2 \rangle^{ \frac{1}{2}} \approx 0.83$ fm. However,
as discussed above, the effect of the pion cloud is expected to increase the root mean square radius
by about a third. The incompressibility appears to be in the generally accepted 
range for the larger values of $\alpha$. The $\approx 20$ \% uncertainty in this quantity 
is again due to a combination of crystal-structure, tolerance and fit-induced uncertainties.
As mentioned above, the experimental value of the pion decay constant,
$f_{\pi} = 93$ MeV is best reproduced in our simple single-parameter description
for $\alpha =1.05$ GeV. The value of $f_{\pi}$
is important for calculations in the hadronic sector. It is anticipated that 
with a more complicated gluon propagator with several parameters, the quantities
$f_{\pi}$, $\rho_{min}$ and $K$ can all have reasonable values simultaneously.

\begin{figure}
\vspace*{8cm}
\includegraphics{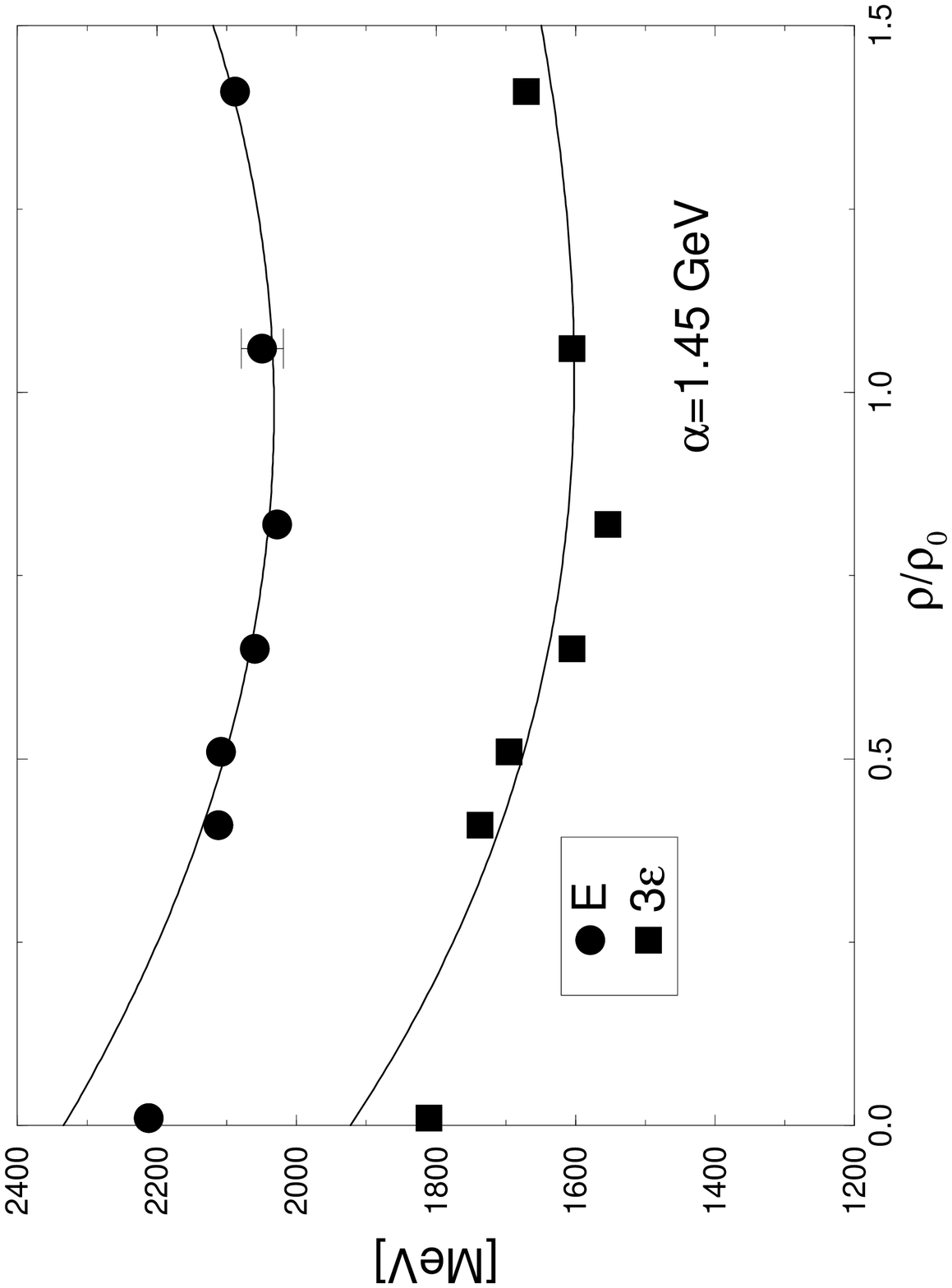}
\caption[]{The total energy $E$ and the quark contribution to the total energy in MeV, 
as a function of $\rho/\rho_0$ for $\alpha=1.45$~GeV.}
\label{fig3}
\end{figure}

\section{Discussion}\label{conc}

We have calculated saturation properties of nuclear matter with a nonlocal confining 
soliton model, the Global Color Model (GCM). The encouraging successes of the GCM
on the hadronic level triggered the use of the model for the description of nuclear
matter in terms of the underlying quark and gluon degrees of freedom. Since there
is a large number of degrees of freedom to deal with in each individual soliton,
the soliton matter picture of nuclear matter is necessarily restricted to the mean
field level and the Wigner-Seitz approximation is utilized. This model provided
robust conclusions on the QGP transition at high density\cite{jff96,jf97}.
We therefore felt it necessary to investigate the properties of the model
at around standard nuclear matter density.

A reasonable reproduction of nuclear matter saturation properties has been found. Some
caution is necessary though when interpreting the physical significance of this agreement, 
since the kinetic energy of the soliton as a whole has been neglected in our calculation,
while the Fermi kinetic energy of the nucleons plays an important role in nuclear
saturation. Another reason while the agreement does not signal a deep physical 
understanding of nuclear matter at the quark level is the fact that we used an
extremely simplified, ``bare-bones'' one-parameter gluon propagator. Furthermore,
explicit pions were not included in the model. We find it
remarkable that such a simple effective description nevertheless gives values of
the equilibrium density and of the incompressibility in the right ball park.
We wish to inform the community about the existence of such a powerful effective
description.
\begin{table}
\caption{Variations of the pion decay constant, the root mean square radius of 
a single soliton, the density where the total energy $E$ is minimum, and the incompressibility 
with the single model parameter $\alpha$.} 
\begin{center}
\renewcommand{\footnoterule}{\kern -3pt} 
\begin{tabular}{||l||c|c|c||}   
$\alpha$ (GeV) 			  & 1.25 & 1.35 & 1.45 \\  \hline
$f_{\pi}$ (MeV)                   & 111  & 120  & 129  \\
$\langle r^2 \rangle^{1/2}$ (fm)  & .71  &  .67 &  .64 \\ 
$\rho_{min}/\rho_0$ & $0.66 \pm 0.10$  & $0.73 \pm 0.11 $ & $ 0.98 \pm 0.15 $ \\
K (MeV) & $137 \pm 30$ & $220 \pm 40$  & $309 \pm 50 $ 
\end{tabular}
\label{t1}
\renewcommand{\footnoterule}{\kern-3pt \hrule width .4\columnwidth 
\kern 2.6pt} 			
\end{center}
\end{table}
It is of interest to calculate in-medium properties as a function of the density
with the model. We made the first steps in this direction elsewhere\cite{johnson98}
by calculating the axial-vector coupling constant and correlation function in medium.
It should be kept in mind that more realistic gluon propagators are now available 
in the literature\cite{tandy97}, and should improve the model and accommodate a 
simultaneous fit of the pion decay constant and the root mean square radius 
of the proton. A more complete model requires
the treatment of explicit pion degrees of freedom for inclusion into the calculation 
of the root mean square proton radius. Developments along these lines will also 
facilitate the exposure of the chiral content of the model and direct comparisons
to QCD-based calculations. With such improvements the model can become a practical
tool in the description of high-density strongly interacting matter.

\section{Acknowledgement}

This work was supported in part by the 
U.S. Department of Energy under Grant No. DOE/DE-FG02-86ER-40251.

\section*{Notes} 
\begin{itemize}
\item[ *] Dedicated to the memory of Vladimir Naumovich Gribov.
\item[ \dag] E-mail: johnson@ksuvxd.kent.edu
\item[ \ddag] E-mail: fai@ksuvxd.kent.edu
\item[ \S] We use the convention $\hbar=c=1$ thruoghout.   
\end{itemize}

\end{document}